\documentstyle[times,graphics,astrobib,amssymb]{mn2e}

\newcommand{\be}{\begin{equation}}
\newcommand{\e}{\end{equation}}
\newcommand{\bear}{\begin{eqnarray}}
\newcommand{\ear}{\end{eqnarray}}
\newcommand{\nline}{\nonumber \\}
\newcommand{\f}{\frac}
\newcommand{\de}{{\rm d}}

\title[Cosmology with GRBs]
{Probing the dark ages with redshift distribution of GRBs}
\author[Choudhury \& Srianand] 
{T. Roy Choudhury\thanks{E-mail: tirth@iucaa.ernet.in},
        R. Srianand\thanks{E-mail: anand@iucaa.ernet.in}\\
        IUCAA, Post Bag 4, Ganeshkhind, Pune 411 007, India.}

\begin{document}

\maketitle 
\begin{abstract}
In this article, we explore the possibility of using the properties of gamma 
ray bursts (GRBs) to probe the physical conditions in the epochs prior
to reionization. 
The redshift distribution of GRBs is modelled 
using the Press-Schechter formalism with an 
assumption that they follow the cosmic
star formation history. 
We reproduce the observed star formation rate 
obtained from 
galaxies in the redshift range $0 < z < 5$, 
as well as the redshift distribution 
of the GRBs inferred from the luminosity-variability 
correlation of the 
burst light curve. 
We show that the fraction of GRBs at high redshifts,  
whose afterglows cannot be observed in 
R and I band
due to H{\sc i} Gunn Peterson optical depth can, at the most,  
account for \emph {one third of the dark GRBs}.
The observed redshift distribution of GRBs, 
with much less scatter than the one 
available today, can put stringent constraints on the epoch 
of reionization and the nature of gas cooling in the epochs prior 
to reionization.
\end{abstract}

 
\section{Introduction} 

The usefulness of gamma ray bursts (GRBs) to probe the evolution
of the universe is realized ever since the redshift measurements
of GRBs became possible \cite{kdr++98}. A direct connection 
between the GRBs and  star formation rate (SFR) in the host galaxy
is expected in the two popular models for GRB progenitors 
(collapsar and binary coalescence). There are also growing
observational indications for the association of supernova-like 
components in several afterglows \cite{gbw++99,bkd++99}. If indeed 
the rate of formation of GRBs is related to the SFR 
of host galaxy, they can be used as an effective probe of the star 
formation history of the universe \cite{lr00,pm01,cl00,bn00,bl02}. 
Thus the redshift distribution of GRBs can complement the studies
based on Lyman break galaxies in understanding the cosmic star formation
history. At present only few tens of GRBs have redshift measurement
and the list is, by no means, statistically complete. Recently
it has been suggested that the redshift distribution of the GRBs
can be probed using the global correlation that seems to exist
between different observables of the afterglow. The redshift
distribution of GRBs, albeit with large scatter, can be obtained using 
the correlation between GRB luminosity and variability 
\cite{fr00,rlfrch01,lfr02}. 

The formation of galaxies and cosmic star formation histories have been 
studied extensively through analytical calculations and numerical 
simulations (see, for example, \citeN{ro77}, \citeN{wr78},
\citeN{kwh96}, \citeN{go97}, \citeN{co00}). 
These studies have been extended to investigate 
the cosmological implications of the GRBs. In this work, we take a simple 
model of the cosmic star formation history, based on Press-Schechter 
theory of collapsing 
haloes 
(in this work, the term `halo' is used for 
referring to the dark matter halo)
and elementary ideas on cooling and galaxy 
formation. The SFR density at low redshift in the model
is constrained by the rest-UV luminosity observations of galaxies 
\cite{temdb98,csb99,sagdp99,spf01}. 
However, at high redshifts (say $z \ge 6$) the nature of reionization
(viz. the epoch of reionization and nature of baryonic cooling in
haloes) strongly influence the star formation history. The main aim
of this article is to use observed redshift distribution of GRBs
to probe the physical processes in the pre-reionization epoch.
Indeed, one cannot probe this epoch directly in the optical 
wave band due to high Gunn-Peterson
optical depth even if the intergalactic medium (IGM) 
is not completely neutral.

\section{Analytical Formalism and Model Parameters}

We use the Press-Schechter (PS) formalism to obtain the comoving number 
density of collapsed objects having mass in the range $(M, M + \de M)$,
which are formed at the redshift interval $(z_c, z_c + \de z_c)$ 
and observed at redshift $z$ \cite{sasaki94,co00}
\bear
N(M,z,z_c) \de M \de z_c &=& 
N_M(z_c) \left(\f{\delta_c}{D(z) \sigma(M)}\right)^2 
\f{\dot{D}(z_c)}{D(z_c)} \nline
&\times&\f{D(z_c)}{D(z)} \f{\de z_c}{H(z_c) 
(1 + z_c)} \de M
\ear
where the overdot represents derivative with respect to time.
The parameter
$N_M(z_c) \de M$ is the number of collapsed objects per unit comoving 
volume within a mass range $(M, M+\de M)$ at redshift $z_c$, known as 
the PS mass function \cite{ps74}, and $\delta_c$ is the 
critical overdensity for collapse, usually taken to be equal to 1.69 
for a matter dominated flat universe $(\Omega_m=1)$. 
This parameter is quite insensitive to cosmology and hence the same value 
can be used for all cosmological models \cite{ecf96}. The other parameters 
are: $H(z)$ is the Hubble parameter, 
$D(z)$ is the growth factor for linear perturbations and  
$\sigma(M)$ is the rms mass fluctuation at a mass scale~$M$.

At early epochs, (i.e., redshifts larger than the reionization
redshift, $z > z_{\rm re}$), the lower mass cutoff for the 
haloes which can host star formation will be
decided by the cooling efficiency of the baryons. Indeed, the absence of 
heavier elements and lower freezout molecular fraction of hydrogen
($f_{{\rm H}_2} \simeq 10^{-6}$) in the primordial gas favours pure atomic
cooling of the gas. This process can only make the gas cool upto 10$^4$ K. 
However, if one can somehow increase  H$_2$ (and probably HD) content of
the gas, the final temperature can be reduced which can lead to the
formation of 
cold gas condensations within the 
low mass haloes \cite{tsrbap97}. 
Density enhancement during the collapse of the  haloes can 
increase the molecular fraction upto $10^{-4}$ in central parts of the 
haloes \cite{har00}. However, the Lyman and Werner band photons
that are produced by luminous objects can easily destroy
these H$_2$ molecules. \citeN{har00} have shown that 
the  haloes smaller than $T_{\rm vir} = 10^{2.4}$ K cannot cool 
even in the absence of any Lyman and Werner band flux. 
In what follows we consider models with $T_{\rm vir} = 300$,
$10^3$ and $10^4$~K separately, which 
essentially covers 
the whole range of possibilities.

After the universe has been reionized ($z < z_{\rm re}$), 
the lower mass cutoff for the  haloes which are 
able to host star formation is set by the 
photoionization temperature ($T \simeq 10^4$ K).  
It is known from simulations that the photoionizing
background suppresses galaxy formation within  haloes 
with circular velocities less than 35 km s$^{-1}$, while the mass 
of cooled baryons is reduced by 50\% 
for  haloes with circular velocities $\sim$ 50  km s$^{-1}$ 
\cite{tw96}. However, the exact value will depend on the 
intensity and spectral shape of the ionizing background radiation.
We consider this as a free parameter spanning a range of 35 
km s$^{-1}$ to 75 km s$^{-1}$ for the circular velocity cutoff 
($v_c^{\rm cut}$).
Note that $v_c^{\rm cut}$ 
is taken to be time-independent -- hence, it does not correspond to  
the linear theory Jeans mass (which 
decreases with time) \cite{gnedin00}.
Also, in our model, the change from the neutral to reionized phase happens
abruptly at $z = z_{\rm re}$. 
However, in realistic models one needs to consider the 
transition region where considerable fraction of gas remains in the 
ionized phase.
This will lead to an  extra suppression in the low mass haloes even in
the pre-overlapping era. 

We assume that the SFR, of a  halo of mass $M$ which is 
formed at redshift $z_c$, is given by \cite{els62,co92,gnedin96}
\bear
\dot{M}_{\rm SF}(M,z,z_c) &=& \epsilon_{\rm SF} 
\left(\f{\Omega_b}{\Omega_m} M \right) 
\f{t(z)-t(z_c)}{t_{\rm dyn}^2} \nline
&\times&
\exp\left[-\f{t(z)-t(z_c)}{ t_{\rm dyn}}\right]
\ear
where $\epsilon_{\rm SF}$ is a efficiency parameter for 
the conversion
of gas into stars. 
The function $t(z)$ gives the age of the universe at redshift $z$; 
thus, $t(z)-t(z_c)$ is the age of the collapsed halo at $z$.
We have assumed here that the duration
of star formation activity in a given  
halo extends over a dynamical 
time-scale $t_{\rm dyn}$.  
The exponential decrease of the SFR comes from the assumption that it 
is proportional to the mass of the cold gas 
(= total baryonic mass {\it minus} the mass already gone into stars)
inside the halo (Schmidt's law).
We use data obtained from rest-UV luminosities of galaxies, 
compiled and corrected for extinction by \citeN{spf01}, 
to normalize $\epsilon_{\rm SF}$. 
Physically, this quantity depends on the metallicity 
of the gas and the effect of feedback due to outflows
and local radiation field.
For simplicity, we assume this to be 
independent of redshift for most of this article. 
Now we can write the cosmic SFR per unit comoving 
volume at a redshift $z$,
\begin{equation}
\dot{\rho}_{\rm SF}(z) = \int_z^{\infty} \de z_c 
\int_M^{\infty} \de M' \dot{M}_{\rm SF}(M',z,z_c) 
\times N(M',z,z_c).
\end{equation}
The lower mass cutoff $M$ at a given epoch is decided by the 
Jeans criteria as explained above. 

Assuming the formation GRBs is closely related to the formation 
of stars, we get the birthrate of GRBs at redshift $z$ 
per unit comoving volume as 
\be
\phi(z) = f \dot{\rho}_{\rm SF}(z)
\e
where $f$ is a efficiency factor which links the formation of
stars to that of GRBs. It is the number of GRBs per 
unit mass of forming stars, hence it depends on the fraction of 
mass contained in (high mass) stars which are potential progenitors 
of the GRBs. This implies that $f$ is crucially 
dependent on the slope of the 
stellar initial mass function (IMF).

Given the formation rate, it is straightforward to obtain the rate 
of GRB events 
per unit redshift range
expected over the whole sky at present:
\be
\f{\de \dot{N}^{\rm GRB}}{\de z} = \phi(z) \f{1}{1+z} \f{\de V(z)}{\de z}
\e
where the factor $(1+z)$ is due to the time dilation between $z$ and the 
present epoch, and $\de V(z)$ is the comoving volume element
\be
\de V(z) = 4 \pi c ~ \f{d_L^2(z)}{H(z) (1+z)^2} \de z
\e
The parameter $d_L(z)$ is the luminosity distance. Note that the 
difference between $\phi(z)$ and $\de \dot{N}^{\rm GRB}/\de z$ is 
purely geometrical in nature. 
The number of GRBs per 
unit redshift range observed over a time $\Delta t_{\rm obs}$ 
is then given by
\be
\f{\de N}{\de z} = \f{\de \Omega}{4 \pi}
\Delta t_{\rm obs} \Phi(z) \f{\de \dot{N}^{\rm GRB}}{\de z}
\label{dndz}
\e
where $\de \Omega/4 \pi$ is the mean 
beaming factor and $\Phi(z)$ is the 
weight factor due to the limited sensitivity of the detector, because of 
which, only brightest bursts will be observed at higher redshifts.

\section{Results and Discussion:}

\begin{center}
\begin{figure*}
\resizebox{0.99\textwidth}{!}{\includegraphics{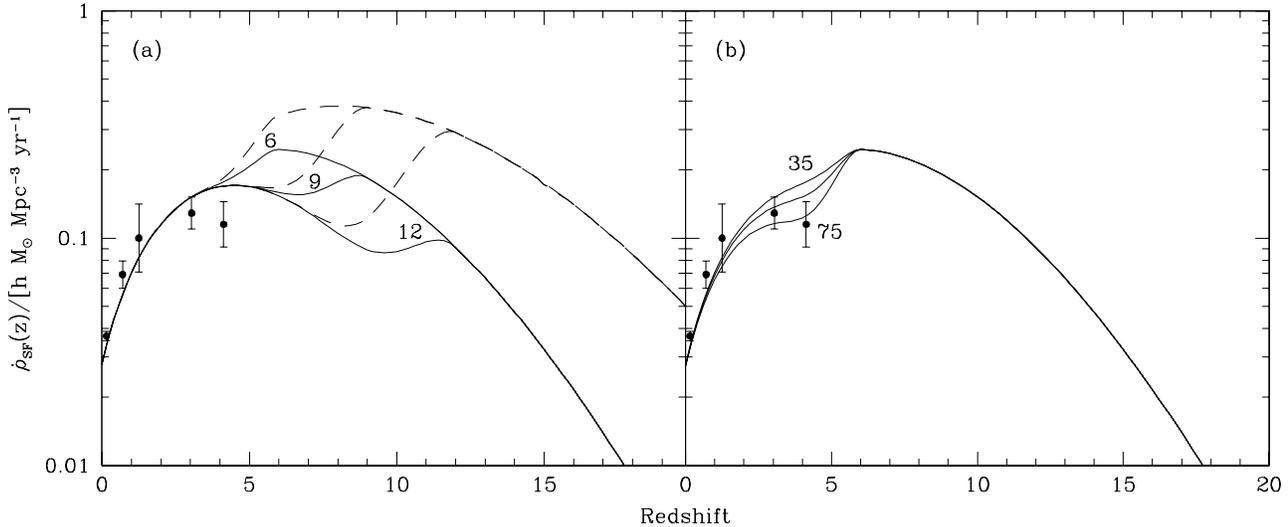}}
\caption{
The SFR density as a function of
redshift. 
In the panel (a) we plot
the results for three epochs of reionization ($z_{\rm re} = 6$, 9 and 12) 
for $T_{\rm vir} = 10^4$ K(solid curves) and 300 K (dashed curves) considering
$v_c^{\rm cut} = 35$ km s$^{-1}$.
In panel (b) the results are plotted for 
$v_c^{\rm cut}$ = 35, 50 and 75 km s$^{-1}$ considering 
$T_{\rm vir} = 10^4$ K and $z_{\rm re} = 6$.
The curves are normalized using 
the extinction corrected 
data points 
from Somerville, Primack, \& Faber (2001).}
\label{fig1}
\end{figure*}
\end{center}

The analysis presented here is based on  LCDM model with the 
parameters $\Omega_m=0.3, \Omega_{\Lambda}=0.7, h=0.65, 
\Omega_b h^2=0.02, \sigma_8=0.93, n=1$.
We consider three epochs of reionization ($z_{\rm re}$),  
three virial temperatures ($T_{\rm vir}$) upto which the gas can cool before
reionization (i.e $z \ge z_{\rm re}$) and three circular
velocity cutoff (${v_c^{\rm cut}}$) for epoch after
reionization. 
For a given set of model parameters,
changing the reionization epoch is equivalent to changing 
the escape fraction of Lyman limit photons from the haloes.
The SFR density obtained from our model is plotted as a function
of redshift in Fig~\ref{fig1}. In the panel (a) we plot
the results for three epochs of reionization for $T_{\rm vir} = 
10^4$ K(solid curves) and 300 K (dashed curves) considering
$v_c^{\rm cut} = 35$ km s$^{-1}$.
\vspace{-0.6cm}

Our models consistently reproduce the observed points 
inferred from 
rest-UV luminosities of 
galaxies at $z \le 5$
for $\epsilon_{\rm SF}$ of 0.1. Indeed the maxima in the curve
occurs in the pre-reionization era. This is mainly due to the
substantial contribution from the low mass haloes where
the star formation is not suppressed by the photoionization 
\cite{bl00,bl02}.
However, as expected, the SFR density depends 
strongly on the epoch of reionization and the nature of the cooling
(hence  $T_{\rm vir}$). 
Interesting point to
note is that across the reionization epoch there is 
upto 3 fold increase in the SFR density. This should
be considered as a strict upper limit as 
is achieved when the Lyman and Werner band photons are completely
absent (hence $T_{\rm vir}=300$ K). 
Additional difference of factor two could be introduced 
across $z_{\rm re}$ by increasing the $v_c^{\rm cut}$ to 
75 km s$^{-1}$, thereby 
suppressing star formation in more low-mass haloes 
in the post-recombination era, which is shown in panel (b).

\subsection{Dark GRBs:}

The optical afterglows are only detected in one third of the well
localized GRBs 
\cite{fjg+01,lcg02,ry01}.
The GRBs without afterglows are called dark GRBs. 
The optical limits obtained in the case 
of dark afterglows are fainter than the detected ones. 
This suggests 
that the dark GRBs can not be accounted for by the failure to image 
deeply enough quickly 
(see \citeN{rp02} and  references therein).
The reason for these missing optical afterglows is attributed 
either (i) to the extinction in the host galaxy, or (ii) to the fact 
that some of the GRBs lie beyond the reionization epoch and the 
neutral IGM absorbs the afterglow. Indeed high-$z$ QSO observations
show complete Gunn-Peterson absorption for $z \ge 6$ QSOs
\cite{fnl+01,dcs+01}.

Our model predicts the redshift distribution of GRBs for a given $f$. 
Realistically, time evolution of the mass function 
of stars (for example triggered by the evolving metallicity) 
and the change in mean properties of the interstellar medium (ISM) 
will introduce time 
dependency for $f$. However, here we assume 
$f$ to be constant.

To predict the {\it observed} redshift distribution, one 
needs to take into account the fact that only the brightest bursts 
will be observed at higher redshifts due to limited detector 
sensitivity.
We take this into account by 
assuming a luminosity function of the form
\be
\psi(L) \propto \left(\f{L}{L_0}\right)^{\gamma} 
\exp\left(-\f{L_0}{L}\right)
\e
and then calculating the fraction of GRBs observed, $\Phi(z)$, at a 
particular $z$ using the relation
\be
\Phi(z) = \f{\int_{L_{\rm min}(z)}^{\infty} \psi(L) \de L}
{\int_0^{\infty} \psi(L) \de L}
\e
Here, 
$L$ is the ``isotropic 
equivalent'' intrinsic burst luminosity in the energy band $30 - 2000$ keV 
(as defined by \citeN{pm01}) and 
$L_{\rm min}(z)$ is the corresponding 
minimum intrinsic luminosity that can be 
observed with the detector.  
We consider two separate cases -- {\bf Case A:} $\psi(L)$, and hence, 
$L_0$ is independent of 
$z$, and {\bf Case B:} $\psi(L)$ evolves with redshift, with the evolution 
given by $L_0 \sim (1+z)^{1.4}$ as suggested by \citeN{lfr02}.

For a given detector sensitivity, 
we can determine $L_{\rm min}(z)$ from the relation
\be
L_{\rm min}(z) = 
P \f{4 \pi d_L^2(z) 
\int_{30~{\rm keV}}^{2000~{\rm keV}} E S(E) \de E} 
{(1+z) 
\int_{(1+z) E_{\rm min}}^{(1+z) E_{\rm max}} S(E) \de E}
\label{burstlum}
\e
where $P$ is the minimum observed photon flux (sensitivity) at the detector 
in the energy band $[E_{\rm min}, E_{\rm max}]$ and $S(E)$ is the 
rest frame GRB energy spectrum. $S(E)$ is taken to be 
a broken powerlaw with 
a low-energy spectral 
index $\alpha$, a high-energy spectral 
index $\beta$ and a break energy $E_b$ \cite{bmf++93}. In this work, 
we take $\alpha = -1$ and $\beta = -2.25$, which are the mean values 
measured by \citeN{pbm++00}, and $E_b = 511$ keV \cite{pm01}.

The minimum intrinsic luminosity at $z$ that can be observed by 
the detector depends on $z$ in two ways: (i) the decrease 
in the flux due 
to distance and (ii) K-correction.
For the 
luminosity function, we take $\gamma = -2.5, 
L_0 = 3.2 \times 10^{51}$ ergs s$^{-1}$ (which are 
obtained by \citeN{pm01} using the $\log N - \log P$ relation 
for BATSE bursts)
and $[E_{\rm min}, E_{\rm max}] = [50, 300]$ keV. We consider
values of $P$ corresponding to two 
detectors, namely, $P = 0.2$ photons cm$^{-2}$ s$^{-1}$ 
(BATSE) 
and $P = 0.04$ photons cm$^{-2}$ s$^{-1}$ ($Swift$) \cite{lr00}.

\begin{figure}
\resizebox{0.49\textwidth}{!}{\includegraphics{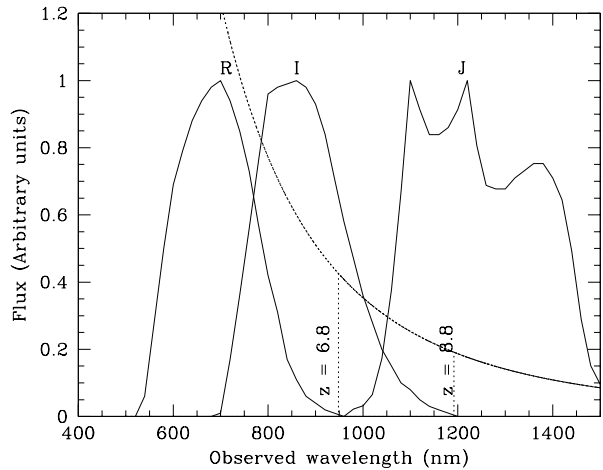}}
\caption{Powerlaw spectrum of GRB afterglow at different
redshifts (labeled at emission wavelength of the Lyman-$\alpha$
line) modified by Gunn-Peterson effect of neutral IGM.
The effect of damping wings is not considered here.
We also plot the window function of the R, I and J bands.
This figure illustrates the fact that afterglow will be 
invisible in R and I band for $z \ge 6.8$ and 8.8 respectively.}
\label{fig2}
\end{figure}

A similar analysis needs to be carried out for the afterglows too. 
According to the simplest afterglow model 
of \citeN{wrm97}, the spectrum follows a power-law form 
when observed in the optical and X-ray wave-bands
about 1 day after the burst. A simple analysis 
shows that the afterglow of the faintest detectable GRB 
will always be brighter than 23rd magnitude within 1 day 
after the burst in both I and R band. 
Hence, if a substantial fraction of GRB afterglows are not observed 
in I and R bands within 1 day after the burst, it must be because of 
radiative transfer effects and \emph {not} because of cosmological effects.

We can now predict the percentage fraction of detectable GRBs at redshifts 
greater than $z$, called $F^{\rm GRB}(z)$. 
From Fig~\ref{fig2}, it is clear that 
due to  high opacity of the IGM neither the afterglows 
nor the corresponding host galaxies will be detected in the R and I-band
if the GRBs are located at 
$z\ge6.8$ and 8.8 respectively. The fraction of R and I band dropout 
afterglows for a range of model parameters and 
detector sensitivities are given in the fourth 
to the seventh 
columns of the Table~\ref{table1} ($F^{\rm GRB}{\rm (R)}$ and 
$F^{\rm GRB}{\rm (I)}$).
The results clearly indicate that even in the optimistic 
case (with luminosity evolution and high sensitivity of the 
detectors), only 38\% (25\%)of the afterglows will not be detected 
in the R (I) band because of the extinction due to the IGM opacity.
It is interesting to note that even
if the H{\sc i} optical depth of the ISM in the host galaxy
is as high as 10$^{23}$ cm$^{-2}$, due to the redshift effect,
X-ray absorption produced by this gas will not affect our 
visibility of this source in the soft X-ray band (0.5 to 2 keV) 
whenever it is detected in the hard X-rays. However such a
gas will produce damped Lyman-$\alpha$ absorption.

\subsection{Redshift distribution of GRBs:}

We next compare the differential redshift distribution of
GRBs 
with that obtained 
using the luminosity - variability $(L - V)$ correlation of GRBs 
\cite{lfr02,sdb01}. 
The variability $V$ 
of a GRB gives a measure of the rms fluctuation 
of the 
burst light curve around the mean value. 
For the GRBs with known redshifts, it was seen that 
there is a correlation 
between the luminosity and the variability of the light curve, 
of the form $L \propto V^a$, 
with $a = 3.35^{+2.45}_{-1.15}$ \cite{fr00}, 
$3.3^{+1.1}_{-0.9}$ \cite{rlfrch01} and 
$1.57^{+0.48}_{-0.54}$ \cite{lfr02}. 
The difference between various results is because of 
the precise definition of variability used.
Since the 
number of GRBs with known redshifts is small ($\sim 20$), 
the scatter 
in the $(L - V)$ correlation is very large. 

\begin{table}
\caption{The predicted range in the percentage 
fraction of dark GRBs due to
Gunn-Peterson effect, taking into account 
two different models for the luminosity evolution (Case A and B). 
The lower and upper limits of $F$ are obtained with 
detector sensitivities similar to that of BATSE and $Swift$ 
respectively.}
\centerline
{
\begin{tabular}{ccccccccc}
\hline
{$z_{\rm re}$} & {$v_c^{\rm cut}$} 
& {$T_{\rm vir}$}&
\multicolumn{2}{|c|}{$F^{\rm GRB}{\rm (R)}$} &
\multicolumn{2}{|c|}{$F^{\rm GRB}{\rm (I)}$} 
\\
&km s$^{-1}$ & K & 
{Case A} & 
{Case B} &
{Case A} & 
{Case B} 
\\
\hline
6  & 75 & $10^4$ & $0.8\!-\!3.9$ & $15.5\!-\!25.6$ 
& $0.2\!-\!0.9$ & $5.6\!-\!11.4$  \\
6  & 50 & $10^4$ & $0.8\!-\!3.5$ & $14.1\!-\!23.7$ 
 & $0.2\!-\!0.8$& $5.1\!-\!10.5$ \\
6  & 35 & $10^4$ & $0.7\!-\!3.3$ & $13.3\!-\!22.5$ 
 & $0.1\!-\!0.7$& $4.8\!-\!10.0$ \\
6  & 35 & $10^3$ & $1.1\!-\!5.0$ & $20.0\!-\!32.9$ 
& $0.3\!-\!1.4$& $8.8\!-\!17.7$  \\ 
6  & 35 & $300 $ & $1.3\!-\!5.7$ & $22.7\!-\!36.9$ 
& $0.4\!-\!1.7$& $10.7\!-\!21.1$ \\
\\
9  & 35 & $10^4$ & $0.6\!-\!2.7$ & $12.3\!-\!21.8$ 
& $0.1\!-\!0.7$& $5.2\!-\!11.1$  \\
9  & 35 & $10^3$ & $0.9\!-\!4.1$ & $19.1\!-\!33.3$ 
& $0.3\!-\!1.5$& $10.1\!-\!20.7$ \\ 
9  & 35 & $300 $ & $1.0\!-\!4.7$ & $22.1\!-\!37.9$ 
& $0.4\!-\!1.8$& $12.4\!-\!24.9$ \\
\\
12 & 35 & $10^4$ & $0.4\!-\!1.9$ & $8.7\!-\! 16.1$ 
& $0.0\!-\!0.5$& $3.7\!-\!8.4 $  \\
12 & 35 & $10^3$& $0.5\!-\!2.3$  & $12.3\!-\!24.0$ 
& $0.2\!-\!0.9$& $7.3\!-\!16.7$ \\ 
12 & 35 & $300 $ & $0.5\!-\!2.5$ & $14.3\!-\!28.0$ 
& $0.2\!-\!1.1$& $9.2\!-\!20.9$  \\
\hline
\end{tabular}
}
\label{table1}
\end{table}

From
this observed correlation, one can, in principle, obtain the 
the redshift distribution of GRBs. 
\citeN{lfr02} use a sample of 220 bursts to obtain the 
intrinsic cumulative 
redshift distribution of GRBs 
$N_{\rm obs}(<1+z)$. Their analysis, that corrects for 
the truncation effects in the sample due to limited 
detector sensitivities, gives the \emph {intrinsic} redshift  
distribution of the GRBs.
We distributed this data into logarithmic redshift bins and calculated 
the slope [which is essentially the 
differential distribution $\de N_{\rm obs}/\de z$] for each bin. 
The error in each bin is mainly contributed by the scatter in 
the $L - V$ correlation.

\begin{figure}
\resizebox{0.49\textwidth}{!}{\includegraphics{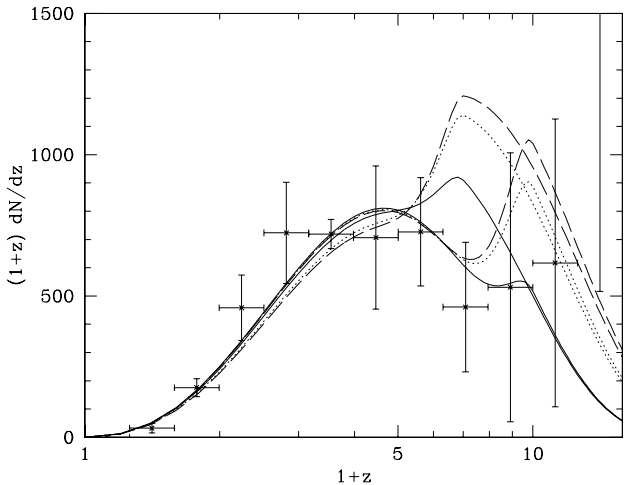}}
\caption{
The redshift distribution of GRBs for different model parameters. 
The data points 
with error bars are from the observed $L -V$ correlation 
(Lloyd-Ronning, Fryer, and Ramirez-Ruiz 2002). 
The error bars along the y-axis represent the 
1$\sigma$ error due to the scatter 
in the observed 
$L - V$ correlation, while the error bars along the x-axis show the bin size. 
The results from our model are shown for $z_{\rm re} = 6$ and 9, and for 
three different values of $T_{\rm vir}$ -- namely, 
$10^4$ K (solid curves), $10^3$ K (dotted curves) and 300 K (dashed curves).
}
\label{fig3}
\end{figure}

The quantity predicted by our model $\de N/\de z$ 
(given by eq (\ref{dndz}), but without the factor $\Phi(z)$ 
in the right hand side)
contains a
normalization factor $f \Delta t_{\rm obs} \de \Omega/4 \pi$, which is 
fixed by fitting to the data 
using $\chi^2$-minimization technique. 
In Fig~\ref{fig3}, we show 
the comparison between our predictions 
and the data obtained by \citeN{lfr02} 
using the 
$L - V$ correlation.
Most of our models are 
broadly consistent with the data mainly because of large error bars.
It is interesting to note that models 
with $T_{\rm vir} \le 10^3$ K deviate a lot from the mean 
observed points; but large errors, mainly from the large 
scatter in the $L - V$ relation, do not 
exclude these models. If the errors in the observed distribution 
reduce, one will be in a position to constrain the epoch of 
reionization and the contribution of molecular cooling 
(and other feedback mechanisms) in the formation 
of first generation of galaxies.

\subsection{Discussions and Conclusions}

In this article we have modelled the redshift distribution of GRBs 
assuming that they follow the cosmic star formation history. 
The main conclusions are summarized below:

\noindent
(i) A peak in the SFR before reionization is seen because of 
substantial contribution from the low mass haloes where
the star formation is not suppressed by the photoionization
(also noticed by \citeN{bl00} and \citeN{bl02}). We show that the 
redshift distribution of the SFR density 
depends on the reionization epoch and the 
minimum virial temperature ($T_{\rm vir}$) 
of star forming haloes. The minimum virial temperature is set by the 
nature of the cooling -- if H$_2$ cooling is effective, haloes 
with much lower virial temperatures can participate in the star formation.

\noindent
(ii) Our results clearly indicate that even in the optimistic 
case (with luminosity evolution and high sensitivity of the 
detectors), only 38\% (25\%)of the afterglows will not be detected 
in the R (I) band because of the extinction due to the IGM opacity.
This means that a 
substantial fraction of optically dark GRBs ($\gtrsim 66$\%) 
originate because of effects such as dust extinction
\cite{wd00,rtb02,bkb++02}.

\noindent
(iii) Although our predictions vary a lot as we change our model 
parameters, we cannot constrain the parameter space effectively 
because of large scatter in observations. One needs an 
improved observed redshift distribution to make real 
progress in understanding the physical processes 
in epochs prior to the reionization.

\noindent
(iv) In general, 
the efficiency parameters $\epsilon_{\rm SF}$ and $f$ depend
of redshift. The star 
forming efficiency $\epsilon_{\rm SF}$ evolves 
because of its dependence on
the metallicity of the collapsing gas and the effect of feedback due to 
outflows and local radiation field.
Similarly, the parameter $f$ evolves 
as the stellar IMF becomes less top-heavy 
with decreasing $z$ \cite{larson98}.
With improved observational data, 
one can use our model to constrain the evolution of 
these parameters.

One would naively expect that the 
presence of damping wing of H{\sc i} absorption could be useful for
probing the neutral IGM. However, in a simple picture of GRBs 
following supernovae, the presence of H{\sc ii} region 
and damped Lyman-$\alpha$ absorption because of the host galaxy 
will complicate the matter. Thus, we believe that the redshift 
distribution with less scatter may be a better way to 
probe the reionization.

\vspace{0.3cm}
We thank N. M. Lloyd-Ronning for providing us with data 
on the cumulative redshift distribution of GRBs. 
We also thank T. Padmanabhan for useful comments.
T.R.C. is supported by the University Grants Commission, India.

\bibliography{mnrasmnemonic,astropap}
 
\bibliographystyle{mnras}
\end{document}